\documentclass[twocolumn,showkeys,showpacs,preprintnumbers,prd,superscriptaddress,nofootinbib]{revtex4-1}
\usepackage{graphicx}
\usepackage{epsf}
\usepackage{bm}
\usepackage{amsmath}
\usepackage{amsfonts}
\usepackage{amssymb}
\usepackage{epstopdf}
\usepackage{natbib}
\usepackage{hyperref}
\usepackage{xcolor}
\usepackage{verbatim}
\usepackage{multirow}
\usepackage{bm}
\usepackage{hyperref}
\usepackage{float}
\usepackage{booktabs}
 
\definecolor{darkblue}{rgb}{0.0, 0.0, 0.55}
\definecolor{darkred}{rgb}{0.55, 0.0, 0.0}

\usepackage{hyperref}
\hypersetup{
    colorlinks=true, % false: boxed links; true: colored links
    linkcolor=darkblue,
    citecolor=darkblue,
    urlcolor=darkblue}
    
\makeatletter\let\expandableinput\@@input\makeatother

\begin{document}

%\title{Testing Late-Time AdS–dS Transitions: Updated Observational Bounds circa 2026}

\title{A mechanistic modeling framework to interpret ACTH stimulation tests across HPA axis adaptation states and glucocorticoid feedback dynamics}

\author{Mamta Yadav}
\email{mamta211935@cuh.ac.in}
\affiliation{Department of Mathematics, School of Basic Sciences, Central University of Haryana, 123029, India}

\author{Phool Singh}
\email{phoolsingh@cuh.ac.in}
\affiliation{Department of Applied Sciences and Humanities, School of Engineering and Technology, Central University of Haryana, 123029, India}

\begin{abstract}
The hypothalamic pituitary adrenal (HPA) axis is a key regulatory system coordinating endocrine responses to physiological and psychological stress. While the ACTH stimulation test remains a cornerstone of adrenal function assessment, its interpretation is complicated by the dynamic and adaptive nature of the HPA axis under chronic stress exposure. In particular, prolonged stress induces glandular remodeling, glucocorticoid receptor (GR) resistance and delayed feedback recovery, all of which may alter test outcomes without indicating primary adrenal failure. In this study, we present a mechanistic modeling framework that integrates hormonal kinetics, feedback inhibition and functional mass adaptation of the corticotroph and adrenal compartments. We simulate the HPA axis over $180$ days encompassing three phases - baseline, chronic stress and recovery, while introducing a time varying GR resistance function to mimic feedback desensitization and its resolution. Using this framework, we evaluate both low dose ($1 \mu g)$ and high dose ($250 \mu g)$ ACTH stimulation tests across physiological phases. Our simulations show that cortisol responses are highly sensitive to both the magnitude and timing of stress exposure and that ACTH responsiveness is phase dependent and often blunted during recovery due to persistent feedback resistance. Low dose ACTH testing more reliably reflects partial adrenal adaptation, while high dose tests risks masking dysfunction due to supraphysiological drive. These results highlight the limitations of static testing paradigms and suggest that accounting for glandular plasticity and GR feedback dynamics is essential for effective endocrine diagnosis particularly in stress related or treatment induced adrenal disorders.  
\end{abstract}


\maketitle
%\section{Highlights}
%\begin{itemize}
%    \item Mechanistic HPA model integrates circadian, stress, and feedback dynamics.
%    \item Dynamic GR resistance explains hypercortisolism and delayed recovery.
%    \item ACTH stimulation tests reveal phase-dependent adrenal responsiveness.
%    \item Provides systems framework for adrenal diagnostics in stress disorders.
%\end{itemize}
\section{Introduction}
\label{sec:intro}
The hypothalamic pituitary adrenal (HPA) axis is a central neuroendocrine system that governs the body's response to both acute and chronic stress \cite{demorrow2018role}. Through a complex series of hormonal interactions, the HPA axis modulates numerous physiological processes including metabolism, cardiovascular tone, immune response and circadian regulation \cite{sheng2021hypothalamic,nicolaides2014circadian,bose2009stress}. The cascade begins with the release of corticotropin releasing hormone (CRH) in response to stress, stimulating the release of adrenocorticotropin hormone (ACTH) from the pituitary, which in turn promotes cortisol synthesis in the adrenal cortex. Cortisol, the primary glucocorticoid in humans, provides essential feedback inhibition at both the hypothalamic and pituitary levels to maintain systemic homeostasis \cite{walker2022fast, sheng2021hypothalamic, keller2015hypothalamic}. 

Despite its critical role, the HPA axis is highly vulnerable to dysregulation, particularly under conditions of sustained psychological or physiological stress \cite{misiak2020hpa, kinlein2015dysregulated}. Dysfunction in this axis have been implicated in a wide range of disorders including adrenal insufficiency (AI), Cushing's syndrome, chronic fatigue syndrome and major depressive disorder (MDD) \cite{marino2020dysfunction,tomas2013review,tanriverdi2007hypothalamo,keller2017hpa}. In many of these conditions, patients exhibit altered cortisol dynamics, impaired feedback sensitivity and abnormal ACTH or CRH signaling. These pathophysiological changes can persist long after the initial stressor has been removed, contributing to the chronicity and recurrence of these conditions \cite{burcusa2007risk, buckman2018risk,lewinsohn1999first}. 

A key diagnostic tool for evaluating adrenal function is the ACTH stimulation test \cite{whiteman2006acth, ghasham2024predictive}, wherein synthetic ACTH is administered and cortisol output is measured to assess adrenal responsiveness \cite{khare2023adrenocorticotropic, oelkers1996adrenal}. The standard $250 \mu g$ dose has been widely adopted to diagnose primary adrenal insufficiency \cite{ospina2016acth}, while the $1 \mu g$ low dose variant has been proposed to sensitively detect secondary or central adrenal impairment \cite{vaiani2019low}. However, the interpretation of ACTH test results is not always straightforward. Clinical studies have shown that chronic stress, prolonged inflammation or exogenous steroid exposure can alter glandular function and feedback regulation, resulting in blunted or exaggerated cortisol responses independent of true adrenal pathology \cite{hannibal2014chronic,thau2023physiology}. These discrepancies underscore a need to incorporate the physiological context of HPA axis adaptation into diagnostic frameworks.   

Mathematical modeling of HPA axis provides an avenue for deeper mechanistic understanding of its behavior across health and disease states. Existing models in literature have successfully captured key features of hormone secretion, ultradian and circadian rhythms and feedback inhibition using systems of nonlinear ordinary differential equations \cite{yi1999dynamical, jelic2005mathematical, gupta2007inclusion, bairagi2008variability, walker2010origin, vinther2011minimal, sriram2012modeling, andersen2013mathematical}. Among these, a widely studied class of models includes inhibitory Hill functions to represent glucocorticoid mediated feedback and incorporate stress input as a modifiable external signal. However, many models stop short of representing long term glandular adaptations or integrating exogenous ACTH inputs directly into the cortisol production pathway. As a result, they are limited in their ability to simulate clinical testing protocols or to explain the persistent dysregulation observed in conditions like MDD and chronic stress exposure. 

Beyond static measurements, the HPA axis exhibits adaptive dynamics involving structural changes in pituitary corticotroph and adrenal cortex mass \cite{karin2020new}. Under prolonged stress exposure, increased trophic drive from CRH and ACTH leads to adrenal hypertrophy and elevated cortisol output \cite{ulrich2006chronic}, a process modulated by intracellular glucocorticoid receptor (GR) feedback. Resistance to glucocorticoids via downregulation of GR density or signaling efficacy can further disrupt feedback inhibition, perpetuating hypercortisolemia and altering the system's sensitivity to ACTH \cite{oster2017functional}. Notably, the recovery from such states is often protracted as glandular remodeling and receptor re-sensitization unfold over days to weeks. Current diagnostic models do not account for these slow time scale adaptations, leading to challenges in distinguishing functional suppression from irreversible endocrine damage. 

In this work, we build upon a previously established HPA axis model \cite{karin2020new} to improve its ability to clinical diagnostics and long term physiological adaptation. While the original model captures key hormonal dynamics and feedback loops, we introduce several critical extensions to enhance biological realism and translational relevance. Our model integrates hormone kinetics, nonlinear feedback and stress sensitive changes in functional gland mass. We simulate circadian hormone dynamics in a healthy HPA axis, track adaptive responses under prolonged stress and subsequent recovery and evaluate both $1 \mu g$ and $250 \mu g$ ACTH stimulation tests across these physiological phases. Additionally we incorporate time dependent GR resistance to model feedback desensitization and its resolution, thereby capturing the nuanced role of glucocorticoids in endocrine diagnostics. This framework provides a unified systems level approach to interpreting ACTH testing in clinical research settings, offering insight into the temporal dynamics of adrenal function and feedback regulation.  

\section{Methods}
\subsection{Mathematical model}
We base our study on an established non linear mathematical model of the HPA axis, originally developed to capture hormone feedback regulation via differential equations representing CRH ($s_1$), ACTH ($s_2$) and cortisol ($s_3$) dynamics. The original model incorporates hormone secretion, clearance rates, negative feedback mediated by glucocorticoids and long term glandular adaptation (corticotroph and adrenal mass). We extend this framework by incorporating additional components to simulate exogenous ACTH stimulation tests, circadian modulation of stress input and time dependent recovery of glucocorticoid feedback sensitivity. These additions allow for the simulation of both acute clinical testing protocols and chronic adaptation dynamics. 

The extended model is defined by the following system of differential equations \\
\begin{eqnarray}
    \frac{ds_1}{dt}&=&p_1(f_1(s_3,t)u(t)-s_1),\label{1}\\
   \frac{ds_2}{dt}&=&p_2(C(t)f_2(s_3,t)s_1-s_2),\label{2}\\
   \frac{ds_3}{dt}&=&p_3(A(t)(s_2+\omega(t))-s_3),\label{3}\\
   \frac{dC}{dt}&=&p_CC(t)(s_1-1),\label{4}\\
   \frac{dA}{dt}&=&p_AA(t)(s_2-1),\label{5}
\end{eqnarray} 
where $s_1, s_2$ and $s_3$ represent concentrations of CRH, ACTH and cortisol, respectively. $C(t), A(t)$ denote the functional mass of corticotroph and adrenal cells, $p_1, p_2$ and $p_3$ are hormone clearance rates derived from empirical half-lives and estimated by $\frac{\log2}{half-life}$. $p_C$ and $p_A$ are mass turnover rates for corticotroph and adrenal cells, respectively. $u(t)$ represents the stress input modulated by both circadian rhythm and chronic stress. $\omega(t)$ is an exogenous ACTH bolus function. $f_1(s_3,t)$ and $f_2(s_3,t)$ are Hill type inhibitory feedback functions governed by glucocorticoid resistance. Feedback inhibition is modeled using a dynamic Hill function $f(s_3,t)=\frac{1}{1+(\frac{s_3}{p_{GR}(t)})^n}$ with $n=3$ and a time dependent feedback sensitivity $p_{GR}(t)$ that increases under chronic stress and gradually recovers post stress. 
\subsection{Model parameters}
\begin{table}[h!]
\centering
\caption{Model parameters used in HPA axis simulations.}
\begin{tabular}{|c|l|c|c|}
\hline
\textbf{Parameter} & \textbf{Description} & \textbf{Value} & \textbf{Ref.} \\
\hline
$p_1$ & CRH clearance rate               & 0.17 min$^{-1}$ & \cite{karin2020new} \\
$p_2$ & ACTH clearance rate              & 0.035 min$^{-1}$ & \cite{karin2020new} \\
$p_3$ & Cortisol clearance rate          & 0.0086 min$^{-1}$ & \cite{karin2020new} \\
$p_C$ & Corticotroph mass turnover       & 0.099 day$^{-1}$& \cite{karin2020new} \\
$p_A$ & Adrenal mass turnover            & 0.049 day$^{-1}$& \cite{karin2020new} \\
$n$   & Hill coefficient                 & 3     & \cite{karin2020new}\\
\hline
\end{tabular}
\label{tab:params}
\end{table}
We adopted physiological parameter values based on hormone clearance and published models.
\subsection{Dynamic GR feedback}
The glucocorticoid resistance parameter $p_{GR}(t)$ is dynamic, fixed at $5$ during chronic stress and gradually decaying toward $2$ post-stress with an exponential recovery function: \\
\[
p_{\text{GR}}(t) =
\begin{cases}
5, & \text{if } t \leq 1440 \cdot D_{\text{stress}} \\
2 + 3 \cdot e^{-\frac{(t - 1440 \cdot D_{\text{stress}})}{\tau}}, & \text{otherwise}
\end{cases}
\] \\
where $D_{stress} = 30$ days and $\tau = 2000$ minutes controls the feedback recovery rate. \\

\subsection{Stress input and circadian modulation}:\\
To model physiologic circadian rhythm and tonic activation, stress input was defined as: \\
\begin{equation}
    u(t)=u_{base}(t)+0.1(\frac{sin(2\pi t)}{1440})
\end{equation}
where $u_{base} =1.0$ (healthy), $4.0$ (depressed) during stress exposure and $0.2$ after stress. \\
\subsection{Exogenous ACTH injection}
Stimulated ACTH tests were implemented by applying a brief exogenous input $\omega(t)$: \\ 
 \[
\omega(t) =
\begin{cases}
\omega_0, & \text{if } 0 \leq t \leq 5 \text{ minutes} \\
0, & \text{otherwise}
\end{cases}
\] \\
where $\omega =1$ for low-dose tests and $\omega =250$ for high-dose tests. \\ 
\subsection{Numerical implementation}
All the simulations were implemented in Python $3$ using the solve ivp function from scipy.integrate, with the Runge-Kutta $45$ method. Output variables $s_1(t), s_2(t), s_3(t), C(t), A(t)$ were recorded at $1$ minute intervals over the simulation window. 
\section{Results}
\subsection{Circadian hormonal dynamics in the healthy HPA axis}
\begin{figure*}
    \centering
    \includegraphics[width=14cm]{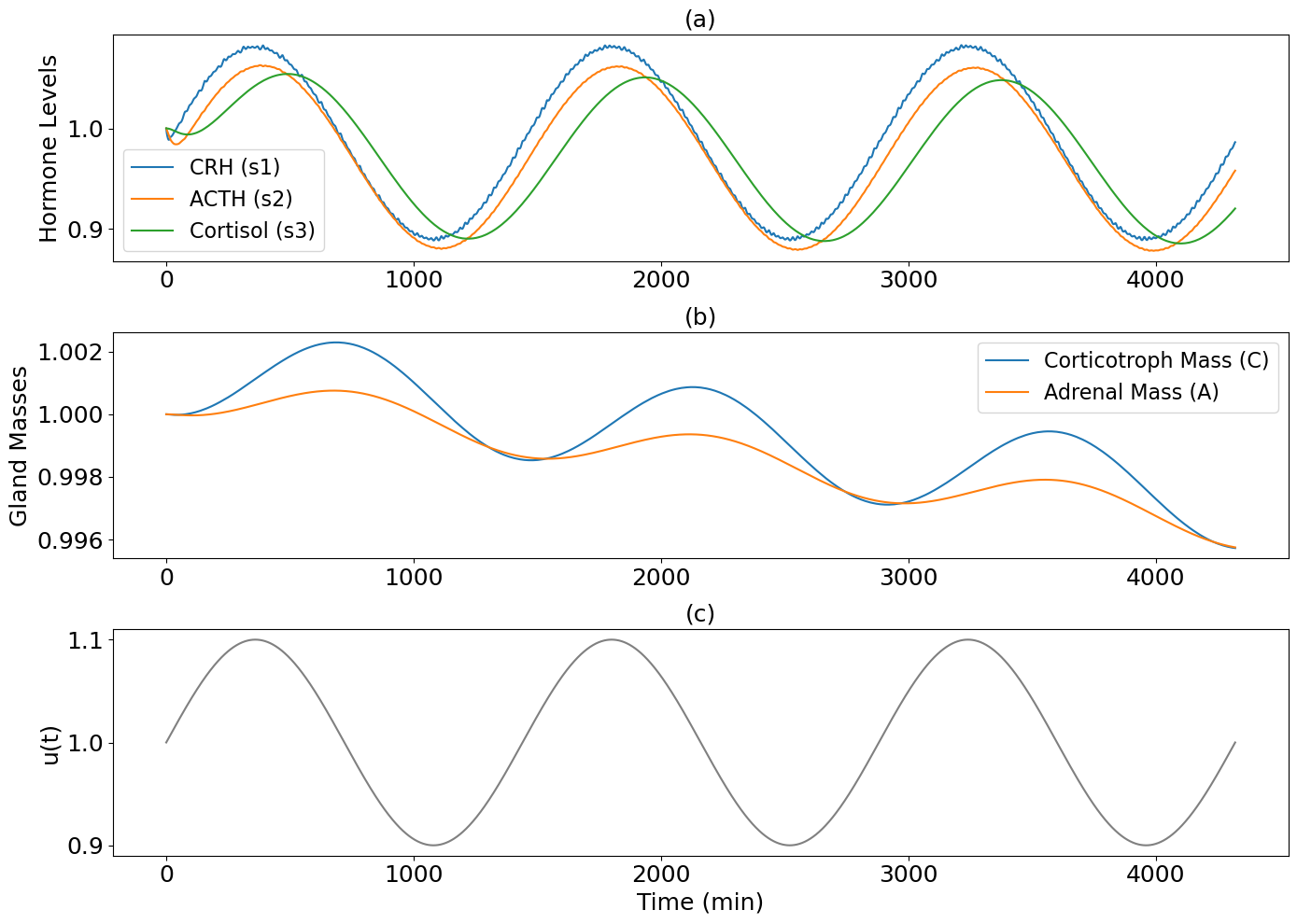}
    \caption{Simulation of the HPA axis over $3$ days under circadian stress input. (a) Hormone levels (CRH ($s_1$), ACTH ($s_2$) and cortisol ($s_3$)) exhibit stable circadian oscillations. (b) Glandular masses (corticotroph and adrenal) remain near baseline, indicating homeostatic stability. (c) Circadian stress input function $u(t)$ driving hormonal dynamics.}
    \label{fig1}
\end{figure*}

To establish a physiological baseline for differentiating adrenal and pituitary dysfunctions, we simulated the full mechanistic HPA axis model under circadian stress input over a $3$ day period. The model incorporated endogenous CRH ($s_1$), ACTH ($s_2$) and cortisol ($s_3$) dynamics, coupled to the functional mass of corticotroph ($C$) and adrenal ($A$) cells. The input stress signal $u(t)$ consisted of a low amplitude sinusoidal function with a $24$ hour period (Figure\ref{fig1}(c)), representing tonic circadian rhythmicity. 

The results demonstrate stable, periodic oscillations in hormone levels, synchronized with the circadian drive. CRH and ACTH exhibit sharp peaks and troughs (Figure\ref{fig1}(a)) in alignment with the sinusoidal stress input, while cortisol displays smoother oscillations due to its longer half life ($80$ minutes) (Figure\ref{fig1}(a)). This delay and dampening of the cortisol waveform are consistent with clinical observations of diurnal cortisol variation. \\
Importantly, both corticotroph and adrenal gland masses remained essentially stable (Figure\ref{fig1}(b)) across the $3$ day simulation window, confirming that under physiological circadian drive, the HPA axis maintains structural and functional homeostasis. These outcomes validate the model's baseline behavior and highlight its capacity to reproduce known features of endocrine regulation, including phase delayed cortisol secretion relative to upstream hormone pulses, negative feedback inhibition from cortisol through non-linear Hill type functions and stability of glandular functional mass in absence of prolonged stress. This healthy circadian behavior serves as a benchmark for contrasting the adaptive responses that arise under pathological conditions. 
\subsection{Adaptive dynamics of the HPA axis under chronic stress and recovery}
\begin{figure*}
    \centering
    \includegraphics[width=14cm]{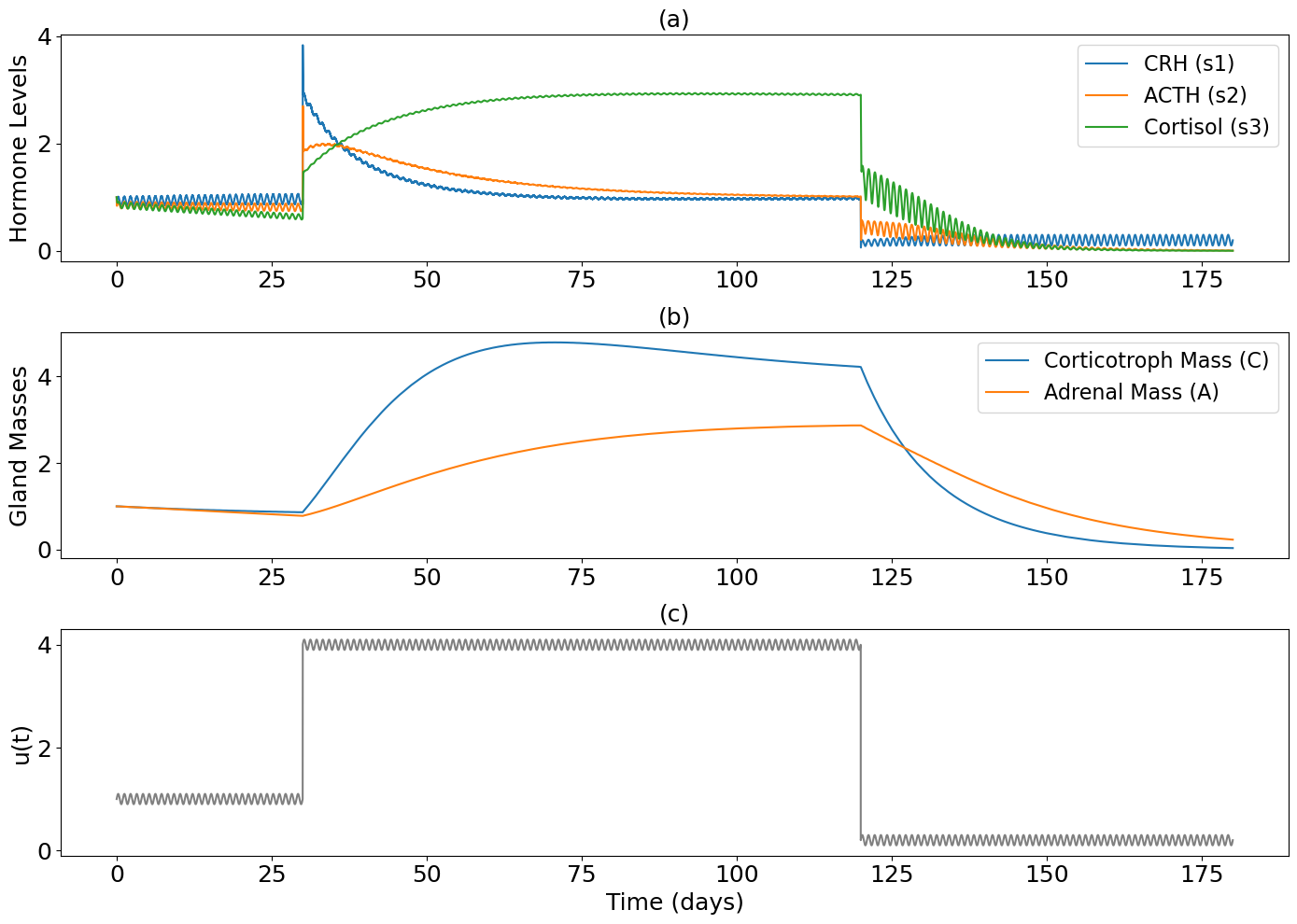}
    \caption{Simulated hormonal and glandular dynamics of the HPA axis across baseline, chronic stress and recovery phases over $180$ days. (a) Hormone trajectories reflect normal circadian variation during the $30$ day baseline phase, elevated activation during $90$ day chronic stress phase and gradual re-stabilization during the $60$ day recovery phase. (b) Corticotroph and adrenal mass increase in response to sustained trophic signaling during stress and exhibit delayed regression upon stress withdrawal, reflecting glandular plasticity. (c) The external stress input $u(t)$, combining phase specific tonic drive and circadian oscillation guides upstream activation.} 
    \label{fig2}
\end{figure*}

To examine the structural and functional plasticity of the HPA axis in response to prolonged stress, we simulated the system over a $30$ day period. The input signal $u(t)$ was composed of a baseline circadian rhythm, a $10$ day phase of chronic stress ($1-10$ days) and a subsequent recovery phase ($11-30$ days) (Figure \ref{fig2}(c)). During the stress period, the tonic stress input was elevated to $u_{base}= 4.0$, followed by a sharp decrease to $u_{base}= 0.2$ (Figure \ref{fig2}(c)) in the recovery phase, simulating a realistic cessation of stress exposure. The results reveal three distinct regulatory phases. In the acute adaptation phase ($1-3$ days), hormone levels (CRH, ACTH and cortisol) rapidly rise in response to elevated $u(t)$, exhibiting amplified circadian oscillations (Figure \ref{fig2}(a)). Corticotroph and adrenal masses begin to increase (Figure \ref{fig2}(b)) driven by sustained elevations in CRH and ACTH, respectively. In the chronic stress phase ($3-10$ days) hormonal outputs plateau at a elevated level, maintaining circadian oscillations but at a higher set point (Figure \ref{fig2}(a)). Functional glandular masses (corticotroph and adrenal) continue to grow (Figure \ref{fig2}(b)), reflecting physiological hypertrophy and the system's effort to maintain homeostasis under high stress. In the recovery phase ($11-30$ days), with the return of a low tonic drive, hormone levels drop sharply (Figure \ref{fig2}(a)), initially dipping below baseline due to inertia in glandular size and negative feedback loop. Gland masses decline gradually but remain above pre-stress levels for several days (Figure \ref{fig2}(b)), indicating slow structural readjustment. Cortisol output normalizes gradually, highlighting the persistence of adaptive changes beyond stress cessation.   
\subsection{$250 \mu g$ ACTH stimulation test across physiological phases}
To evaluate how dynamic adaptation of the HPA axis influences the outcome of ACTH stimulation testing, we employed a mechanistic mathematical model incorporating hormone secretion, clearance, feedback inhibition and glandular plasticity. The system was simulated across three physiologically relevant phases: a homeostatic baseline state (day $30$), a chronic stress state following sustained input (day $118$) and a recovery phase after stress withdrawal (day $150$) (Figure \ref{fig2}(c)). At each time point, an exogenous ACTH pulse of $250$ units administered over $5$ minutes was applied and hormone responses were applied and hormone responses were simulated over a $4$ hour post stimulation window. 

\begin{figure*}
    \centering
    \includegraphics[width=12cm]{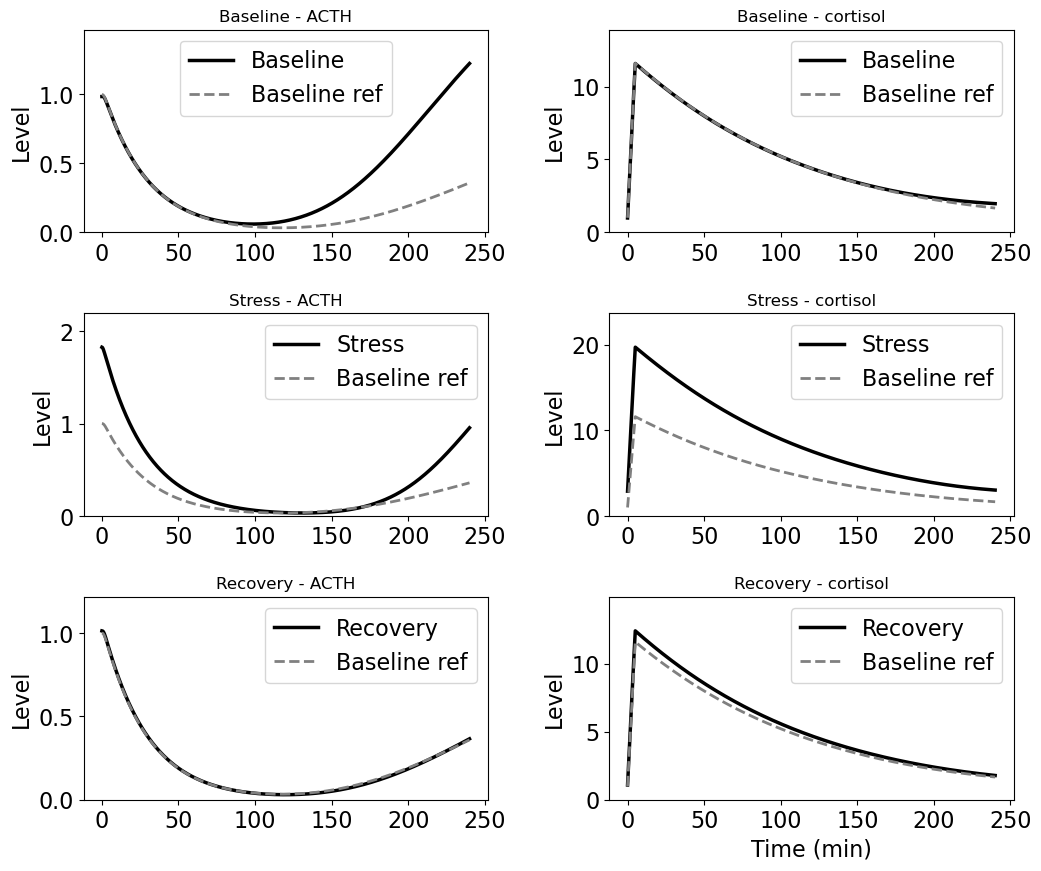}
    \caption{Simulated ACTH and cortisol responses to high dose ($250$ units) ACTH stimulation across three physiological phases. Responses are shown for baseline (day $30$), chronic stress (day $118$) and recovery (day $150$) in a mechanistic HPA axis model incorporating hormone dynamics, feedback inhibition and glandular mass adaptation. A standardized $5$ minute ACTH pulse ($250)$ units was applied at each phase. Cortisol response was attenuated during stress despite increased adrenal mass and partially recovered post stress, highlighting phase dependent variability in test outcomes due to system adaptation. Although absolute cortisol peaks are higher under chronic stress due to elevated baseline secretion, the incremental rise above pre-stimulated levels is markedly reduced, consistent with a blunted dynamic response. (Baseline ref is simply the label given in the legend for a reference ACTH stimulation test performed at the start of the simulation, before any stress or gland size changes occur.) }
    \label{fig3}
\end{figure*}

At baseline, the ACTH stimulation elicited a robust rise in cortisol and ACTH levels, with a peak followed by a gradual return to baseline (Figure \ref{fig3}). This is indicative of a healthy, sensitive adrenal axis, where feedback inhibition via cortisol remains intact and adrenal mass it at steady state functional capacity. In contrast, during the chronic stress phase, the same ACTH pulse resulted in a markedly blunted cortisol response (Figure \ref{fig3}). Despite the adrenal gland being hypertrophied due to sustained trophic stimulation, the responsiveness suppressed, suggesting feedback saturation and possible glucocorticoid receptor desensitization. Endogenous ACTH levels were already elevated due to chronic stress input, reducing the relative impact of exogenous ACTH. The response dynamics during this phase reflect a hyperactivated but a desensitized system, as seen in various stress related disorders.  

By the day $150$ in the post-stress recovery phase, a partial restoration of ACTH and cortisol responsiveness was observed (Figure \ref{fig3}). The cortisol response increased relative to the stress phase but remained attenuated compared to baseline. This lagged recovery is consistent with the slow turnover rates of adrenal and pituitary gland masses and the gradual restoration of GR sensitivity. The results demonstrate that glandular adaptation via mass plasticity and feedback modulation exerts a dominant influence on the interpretation of ACTH stimulation tests. Importantly, the timing of the test relative to the stress exposure and recovery trajectory significantly alters its diagnostic profile. \\

These findings underscore the temporal and physiological state dependence of ACTH test interpretation. The same stimulus yields markedly different outputs depending on the adaptive phase of the HPA axis, suggesting that static thresholds or single time point testing may lead to misclassification of adrenal or pituitary dysfunction. This modeling framework provides a mechanistic rationale for observed clinical variability in stimulation test outcomes and highlights the importance of accounting for system dynamics when using ACTH tests to differentiate between forms of HPA axis dysregulation. 
\subsection{$1 \mu g$ and $250 \mu g$ ACTH stimulation test comparison across physiological phases}
To investigate the diagnostic sensitivity of ACTH stimulation tests under varying physiological conditions, we simulated the HPA axis response to both low dose $1$ unit and high dose $250$ units ACTH challenges at three key time points: baseline (day $30$), chronic stress (day $118$) and recovery (day $150$) (Figure \ref{fig2}(c)). These phases correspond to homeostatic equilibrium, prolonged stress exposure and post stress adaptation, respectively. Using a mechanistic model of the HPA axis that includes hormone kinetics, negative feedback via glucocorticoid signaling and glandular mass adaptation, we simulated $4$ hour hormone response profiles following each ACTH stimulus. 

\begin{figure*}
    \centering
    \includegraphics[width=12cm]{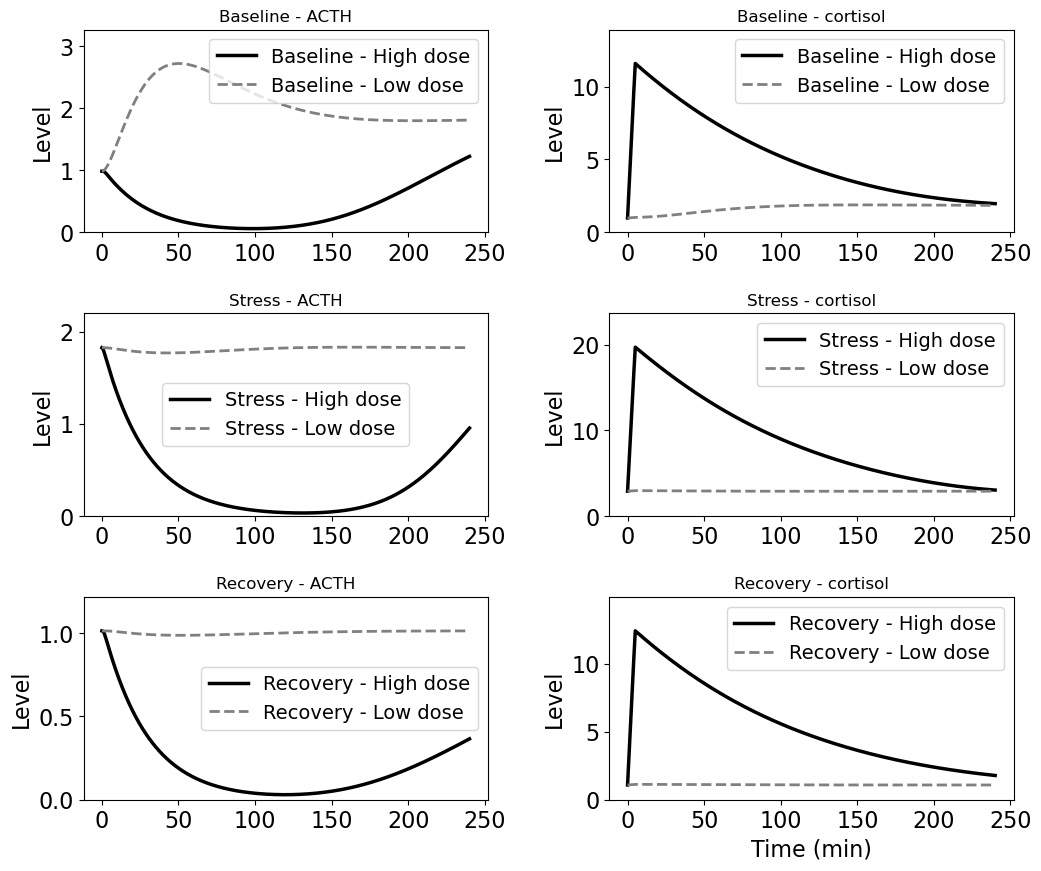}
    \caption{This figure illustrates simulated ACTH and cortisol responses to low- and high-dose ACTH stimulation across baseline, stress, and recovery phases. The model applies $1$ unit (low-dose) and $250$ unit (high-dose) ACTH pulses at three time points: day $30$ (baseline), day $118$ (chronic stress), and day $150$ (post-stress recovery). Cortisol responses are blunted during stress and partially recover by day $150$, while ACTH shows reduced responsiveness under stress and limited sensitivity during recovery. The low-dose test provides finer resolution for detecting adrenal adaptation, especially during recovery. The high-dose cortisol curves have similar shape (fast rise then plateau) because the pharmacodynamics of exogenous ACTH are not phase-dependent in shape, the difference is in magnitude. These results highlight the phase-dependent variability in ACTH test outcomes driven by dynamic HPA axis adaptation.}
    \label{fig4}
\end{figure*}

At baseline, both ACTH and cortisol exhibited strong and well-defined responses to high dose stimulation, with the cortisol curve displaying a rapid rise followed by a smooth decay (Figure \ref{fig4}), reflecting effective adrenal sensitivity and intact feedback control. The low dose ACTH pulse elicited a detectable but reduced response (Figure \ref{fig4}), suggesting that the adrenal system operates below saturation in the healthy state and is capable of scaling output with stimulus intensity. 

Under chronic stress conditions, basal hormone levels were elevated due to sustained upstream input (Figure \ref{fig2}(a)). However, the incremental responses to both low dose and high dose ACTH stimulation were markedly blunted (Figure \ref{fig4}). The cortisol response showed minimal additional elevation beyond the already elevated baseline and ACTH responsiveness was flat, indicating saturation of the feedback loop and possible receptor level desensitization. This supports the model's prediction that during prolonged stress adrenal responsiveness to exogenous ACTH becomes constrained despite increased adrenal mass. These dynamics resemble clinical patterns observed in stress related endocrine dysregulation such as in major depressive disorder. Amsterdam et al. \cite{amsterdam1987acth} tested for cortisol response in laboratory through exogenous administration of ACTH which resulted in elevated cortisol levels in depressed subjects, thus indicating that adrenal cortex may have heightened response to exogenous ACTH. 

In the recovery phase, hormone responsiveness partially rebounded. Cortisol responses to both ACTH doses increased compared to the stress phase but remained below baseline levels (Figure \ref{fig4}), indicating that while glandular mass and receptor sensitivity begin to normalize after stress withdrawal (Figure \ref{fig2}(b)), complete recovery is not instantaneous. Notably, the low dose ACTH test was more sensitive in distinguishing recovery from stress than the high dose test, which may saturate the system and obscure partial functional restoration. 

Together, these simulations illustrate that both the magnitude and shape of the ACTH and cortisol response curves are highly dependent on the system's adaptive state. Importantly, the model demonstrates that low dose ACTH testing may provide greater resolution in assessing partial adrenal responsiveness, particularly during recovery, whereas, high dose tests may underestimate dysfunction due to saturating input.
\subsection{Dynamic GR resistance and its relationship to ACTH regulation and adrenal adaptation}
The glucocorticoids exude inhibitory effect on the HPA axis via negative feedback and chronic inhibition might result in adrenal insufficiency \cite{karangizi2019glucocorticoid,paragliola2017treatment}. Clearly, ACTH secretion may be suppressed by exogenous glucocorticoids and this represents the most common cause of ACTH deficiency. Adrenal insufficiency occurs when the adrenal glands fail to produce enough cortisol. Since the 1950s, it has been recognized that Adrenal Insufficiency can be a side effect of glucocorticoid (GC) therapy \cite{nicolaides2015glucocorticoid, prete2021glucocorticoid, axelrod1976glucocorticoid}. After stopping GC therapy, the Hypothalamic-Pituitary-Adrenal (HPA) axis may remain suppressed, which can lead to AI \cite{axelrod1976glucocorticoid}. Glucocorticoid receptor (GR) signaling plays a key role in managing the negative effects associated with high cortisol levels. When GR feedback is impaired, the HPA axis fails to properly shut down the stress response within a few hours, resulting in excessive cortisol production \cite{karangizi2019glucocorticoid}. Over longer periods, GR helps the HPA axis cope with stress, providing resilience. 

The HPA axis functions as a classic feedback system similar to many other endocrine systems. Glucocorticoids whether naturally produced or externally administered exert negative feedback on the hypothalamus and the pituitary gland. The effects of glucocorticoids can be categorized into two phases: acute and delayed. The acute phase happens within minutes of administration where a rapid rise in cortisol levels suppresses the release of ACTH and CRH \cite{herman2016regulation}. The delayed phase begins between $2$ to $20$ hours after administration and can last for several days. This phase primarily involves the suppression of gene transcription that leads to ACTH production. The intensity of the delayed phase depends on the dose and duration of glucocorticoid use and it becomes more prominent with long term use. While a few doses of glucocorticoid can quickly suppress the HPA axis, recovery is generally fast. However, with prolonged use the recovery of the HPA axis takes much longer \cite{herman2016regulation}. 

\begin{figure*}
    \centering
    \includegraphics[width=12cm]{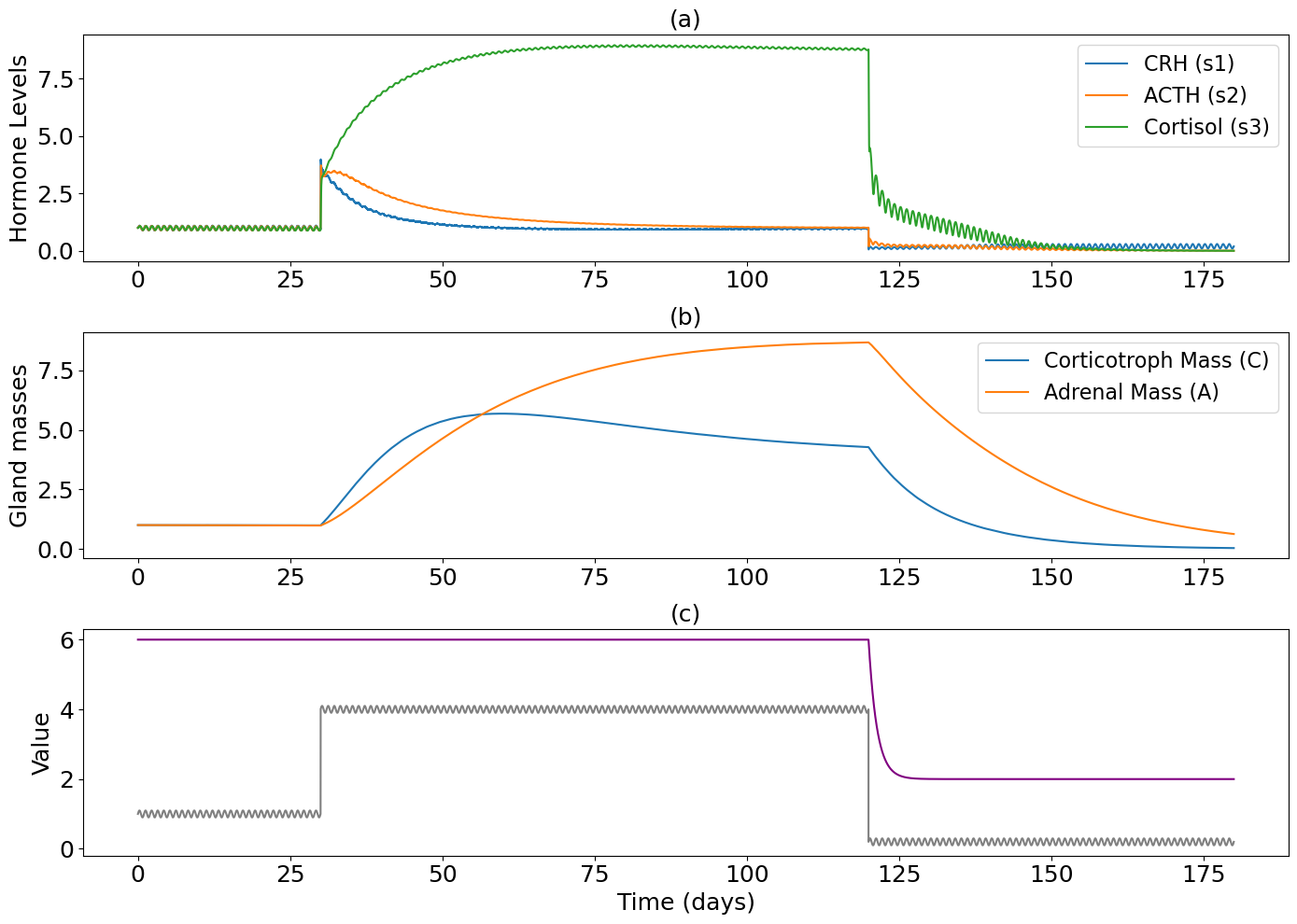}
    \caption{ 
    Simulated dynamics of the HPA axis over $180$ days under variable stress input and time dependent GR resistance. The model captures hormonal and glandular adaptations across three phases: a $30$ day baseline period (low tonic stress), a $90$ day chronic stress period (high input), and a $60$ day recovery period (low stress). (a) CRH, ACTH, and cortisol levels exhibit phase-specific dynamics, with sustained elevation of cortisol and ACTH during stress and gradual normalization during recovery. (b) Functional masses of corticotroph and adrenal glands increase during stress due to trophic signaling and show partial regression in recovery. (c) The stress input function $u(t)$ (gray) reflects circadian variation superimposed on phase-specific baselines, while the GR resistance function 
    $k_{GR}(t)$ (purple) models impaired feedback during chronic stress and exponential recovery thereafter.}
    \label{fig5}
\end{figure*}

This simulation explores the long-term dynamics of the HPA axis in response to varying stress exposure, incorporating a physiologically realistic modulation of GR sensitivity. The model integrates a dynamic resistance parameter $k_{GR}(t)$ which increases during stress and gradually decays post stress to reflect empirically observed alterations in feedback sensitivity (Figure \ref{fig5}(c)). The simulation spans $180$ days and is divided into three distinct phases: an initial $30$ day baseline period, a sustained $90$ day period of elevated stress input and a $60$ day recovery phase characterized by reduced stress drive (Figure \ref{fig5}(c)). Importantly, the model tracks key hormonal (CRH, ACTH, cortisol) and glandular (corticotroph, adrenal) variables over time, all modulated by stress and GR feedback. 

During the chronic stress phase, GR resistance was elevated (modeled by $k_{GR}=6$) (Figure \ref{fig5}(c)), reflecting reduced glucocorticoid sensitivity commonly observed in major depressive disorder, post traumatic stress disorder and critical illness. This impaired feedback leads to sustained elevation of cortisol (Figure \ref{fig5}(a)) and progressive adrenal hypertrophy (Figure \ref{fig5}(b)) consistent with findings in chronic stress exposure. The transition into the recovery phase triggers an exponential normalization of $k_{GR}(t)$, mimicking receptor re-sensitization. Consequently, the model shows a gradual decline in hormone output (Figure \ref{fig5}(a)) and a delayed regression of gland masses (Figure \ref{fig5}(b)), suggesting that recovery of negative feedback precedes morphological remodeling. 

Clinically, ACTH stimulation testing is used to assess adrenal responsiveness and distinguish between primary and secondary adrenal insufficiency \cite{ospina2016acth}. However, test interpretation is confounded in chronic illness due to feedback resistance and HPA axis adaptation. This model highlights that GR resistance though not routinely measured can significantly alter both basal hormone levels and stimulation test outcomes. The simulations emphasize the importance of accounting for GR dynamics in diagnostic interpretation of adrenal function, particularly in stress related endocrine disorders where adaptive changes may mimic or mask pathology.
\section{Discussion}
This study presents a comprehensive mechanistic model of the HPA axis that integrates
circadian hormonal rhythms, stress driven glandular plasticity, dynamic GR feedback and
pharmacological stimulation through ACTH testing. By simulating hormone secretion, tissue
adaptation and feedback resistance over extended time scales, the model offers a systems level
view of the HPA axis physiology and provides critical insights into how dynamic endocrine
states influence the interpretation of diagnostic ACTH stimulation tests.
This study presents a comprehensive mechanistic model of the HPA axis that integrates circadian hormonal rhythms, stress driven glandular plasticity, dynamic GR feedback and pharmacological stimulation through ACTH testing. By simulating hormone secretion, tissue adaptation and feedback resistance over extended time scales, the model offers a systems level view of the HPA axis physiology and provides critical insights into how dynamic endocrine states influence the interpretation of diagnostic ACTH stimulation tests. 

Our results recapitulate well established physiological phenomena such as circadian oscillations in hormone secretion under baseline conditions and adrenal hypertrophy under sustained stress. The simulations demonstrate that chronic stress elevates cortisol and ACTH levels while including mass expansion of corticotroph and adrenal tissue. These adaptations enhance hormonal output but also disrupt homeostatic feedback control. Upon stress withdrawal, hormone levels decline but structural recovery is delayed due to slow turnover rates of endocrine tissues, a feature consistent with histological findings from chronic stress models. 

Importantly, the model incorporates dynamic GR resistance, which is critical in understanding altered HPA feedback during and after stress exposure. Elevated GR resistance during stress promotes hypercortisolemia and blunted ACTH feedback, both of which are common in depression, PTSD and critical illness. Even after stress input is normalized, persistent GR resistance delays hormonal stabilization and alters adrenal responsiveness highlighting a key mechanism behind transient but functionally significant adrenal suppression. Notably, PTSD presents a more complex phenotype where, despite evidence of enhanced GR sensitivity in many patients, cortisol levels are often attenuated rather than elevated. Our framework provides a mechanistic explanation for this apparent paradox; heightened feedback efficiency combined with impaired adrenal reserve or prior structural adaptation can drive hypocortisolism, even as stress-related symptoms persists. A critical insight from our simulations is the role of dynamic GR resistance in shaping both hyper- and hypocortisol states. Elevated GR resistance during stress promotes ACTH hyperdrive, adrenal hypertrophy, and hypercortisolemia. However, after stress cessation, if GR resistance does not rapidly normalize, a paradox emerges: ACTH may rebound, but adrenal cortisol output lags due to prior structural downregulation. This mismatch produces a phase of relative hypocortisolism. Such a pattern provides a mechanistic explanation for the biphasic HPA trajectories often observed in stress-related disorders, ranging from hyperactivity in depression to hypoactivity in subsets of post-traumatic stress disorder (PTSD) patients.

ACTH stimulation tests simulated at multiple phases reveal distinct dynamics. During baseline, both low-dose ($1 \mu g$) and high dose ($250 \mu g$) tests elicit robust cortisol responses. However, during chronic stress, adrenal hyperplasia masks functional suppression in the high-dose test, while the low-dose test more sensitively captures deviations in gland responsiveness. In recovery, residual GR resistance blunts ACTH feedback and prolongs recovery of adrenal sensitivity. These findings align with clinical studies questioning the diagnostic specificity of high-dose ACTH testing in central adrenal insufficiency and support the use of phase aware models for interpreting test results.

Overall, this model offers a new theoretical framework for understanding adrenal diagnostics in dynamic physiological and pathophysiological contexts. It reinforces the idea that ACTH stimulation tests must be interpreted in light of prior stress history, receptor dynamics and glandular plasticity, all of which shape cortisol responses beyond static endocrine deficits. 

While our model captures key features of HPA axis physiology, several limitations must be acknowledged. First, GR resistance function is phenomenological and could be refined with molecular data on receptor turnover, translocation or signaling kinetics. Second, the model assumes homogeneous glandular populations, while in reality, cellular heterogeneity and zonation in the adrenal cortex affect steroidogenesis. Third, it does not yet incorporate other regulatory axes that interact with the HPA system under stress. Additionally, while we model time resolved responses, we don not yet account for stochastic ultradian pulsatility or inter-individual variation.  

Future extensions of this model may incorporate individualized feedback kinetics, multi-axis interactions and real patient data for personalized adrenal function assessment. 
\section{Conclusion}
In this study, we developed a mechanistic, dynamical model of the HPA axis that captures circadian rhythmicity, stress induced hormonal regulation, tissue adaptation and dynamic glucocorticoud feedback. Simulations over $180$ days reveal how chronic stress and delayed recovery reshape both hormone levels and gland structure. By incorporating ACTH stimulation tests within this framework, we demonstrate how diagnostic outcomes vary across physiological phases and are modulated by feedback resistance and adrenal adaptation. Our findings suggest that traditional interpretations of ACTH stimulation tests may be insufficient in contexts involving recent or ongoing stress, where functional adaptation and GR resistance obscure underlying endocrine function. The model highlights the limitations of static dose response diagnostics and emphasizes the importance of time aware systems based approaches in adrenal evaluation.

%\section{Acknowledgements}
%The authors gratefully acknowledge the reviewers for their valuable comments and suggestions, which have significantly contributed to improving the quality of the manuscript.

%\section{Conflict of interest}
%The authors declare that they have no conflict of interest.
%\section{CRediT authorship contribution statement}
%\textbf{Mamta Yadav}: Conceptualization, Methodology, Investigation, Visualization, Writing - original draft. \textbf{Phool Singh}: Supervision, Software, Validation, Writing - review and editing. 
%\section{Funding sources}
%This research did not receive any specific grant from funding agencies in the public, commercial, or not-for-profit sectors.
\section{Appendix}
The governing model of HPA axis \cite{karin2020new} is
\begin{eqnarray}
    \frac{ds_1}{dt}&=&p_1(f_1(s_3)u-s_1),\label{7}\\
   \frac{ds_2}{dt}&=&p_2(Cf_2(s_3)s_1-s_2),\label{8}\\
   \frac{ds_3}{dt}&=&p_3(As_2-s_3),\label{9}\\
   \frac{dC}{dt}&=&p_CC(s_1-1),\label{10}\\
   \frac{dA}{dt}&=&p_AA(s_2-1),\label{11}
\end{eqnarray}
To solve this system, we set $f_1(s_3) = s_3$ and $f_2(s_3) = s_3$, $u$ is the stress input signal which is constant. To obtain the equilibrium points we set the right hand side of each equation equal to zero. So the we get the following results 
\begin{eqnarray}
   p_1(s_3u-s_1)&=&0,\label{12}\\
   p_2(C(s_3)s_1-s_2)&=&0,\label{13}\\
   p_3(As_2-s_3)&=&0,\label{14}\\
   p_CC(s_1-1)&=&0,\label{15}\\
   p_AA(s_2-1)&=&0,\label{16}
\end{eqnarray}
From \eqref{15} and \eqref{16} we obtain 
$C = 0$ or $s_1 = 1$ and $A =0$ or $s_2 = 1$. Clearly $C$ and $A$ represent cell masses and cannot be zero. Consequently $s_1 = 1$ and $s_2 = 1$. We can obtain the remaining correlations by substituting the values of $s_1$ and $s_2$. The equilibrium points are 

    $s_1^*=1,
    s_2^*=1,
    s_3^*=\frac{1}{u},
    C^*=C,
    A^*=\frac{1}{u}$

These values express the equilibrium solutions in terms of the parameters $p_1, p_2, p_3, p_C, p_A$ and $u$. This a steady-state solution based on the system's dynamics. This system of equations is non-linear and solving it analytically is challenging. Instead, we linearize the system around the equilibrium points and find the approximate solution. To linearize this system around the equilibrium points $s_i^*$ we substitute 
\begin{equation}
    s_i=s_i^* + \delta s_i\label{17}
\end{equation}
where $\delta s_i$ is a small perturbation around the equilibrium. After the substitution we expand each differential equation and retain only the first order terms in $\delta s_i$, ignoring the higher order terms. From \eqref{7} we have \\
\begin{equation}
    \frac{d(1 + \delta s_1)}{dt}=p_1((\frac{1}{u} +\delta s_3)u-(1 + \delta s_1))\label{18}
\end{equation}
which gives 
\begin{equation}
    \frac{d(\delta s_1)}{dt}=p_1(u\delta s_3-\delta s_1)\label{19}
\end{equation}
Similarly, by making substitutions in \eqref{8}, \eqref{9}, \eqref{10}, \eqref{11} we get the following linearized system of equations 
\begin{eqnarray}
     \frac{d\delta s_1}{dt}&=&p_1(u\delta s_3-\delta s_1),\label{20}\\
   \frac{d\delta s_2}{dt}&=&p_2(\frac{C}{u}\delta s_1+C\delta s_3+\frac{\delta C}{u}-\delta s_2),\label{21}\\
   \frac{d\delta s_3}{dt}&=&p_3(\frac{\delta s_2}{u} + \delta A -\delta s_3),\label{22}\\
   \frac{d\delta C}{dt}&=&p_CC\delta s_1,\label{23}\\
   \frac{d\delta A}{dt}&=&\frac{p_A\delta s_2}{u},\label{24}
\end{eqnarray}
This system of equations can be written in the matrix form as 
\[
\frac{d}{dt} \begin{bmatrix} \delta s_1 \\ \delta s_2 \\ \delta s_3 \\ \delta C \\ \delta A \end{bmatrix} = A \begin{bmatrix} \delta s_1 \\ \delta s_2 \\ \delta s_3 \\ \delta C \\ \delta A \end{bmatrix}
\] 
where \[
A = \begin{bmatrix}
    -p_1 & 0 & p_1 u & 0 & 0 \\
    \frac{p_2 C}{u} & -p_2 & p_2 C & \frac{p_2}{u} & 0 \\
    0 & \frac{p_3}{u} & -p_3 & 0 & p_3 \\
    p_C C & 0 & 0 & 0 & 0 \\
    0 & \frac{p_A}{u} & 0 & 0 & 0
\end{bmatrix}
\] 
On solving this system we get the following eigen values 
-0.16929341+0.j, 0.03483907+0.j, -0.02052443+0.02614318j, -0.02052443-0.02614318j, -0.02052443-0.02614318j. The eigenvalues reveal that for given $u=4$ (constant) and $p_{GR}=2$ (constant) the system is stable.
\begin{figure}[h]
    \centering
    \includegraphics[width=10cm]{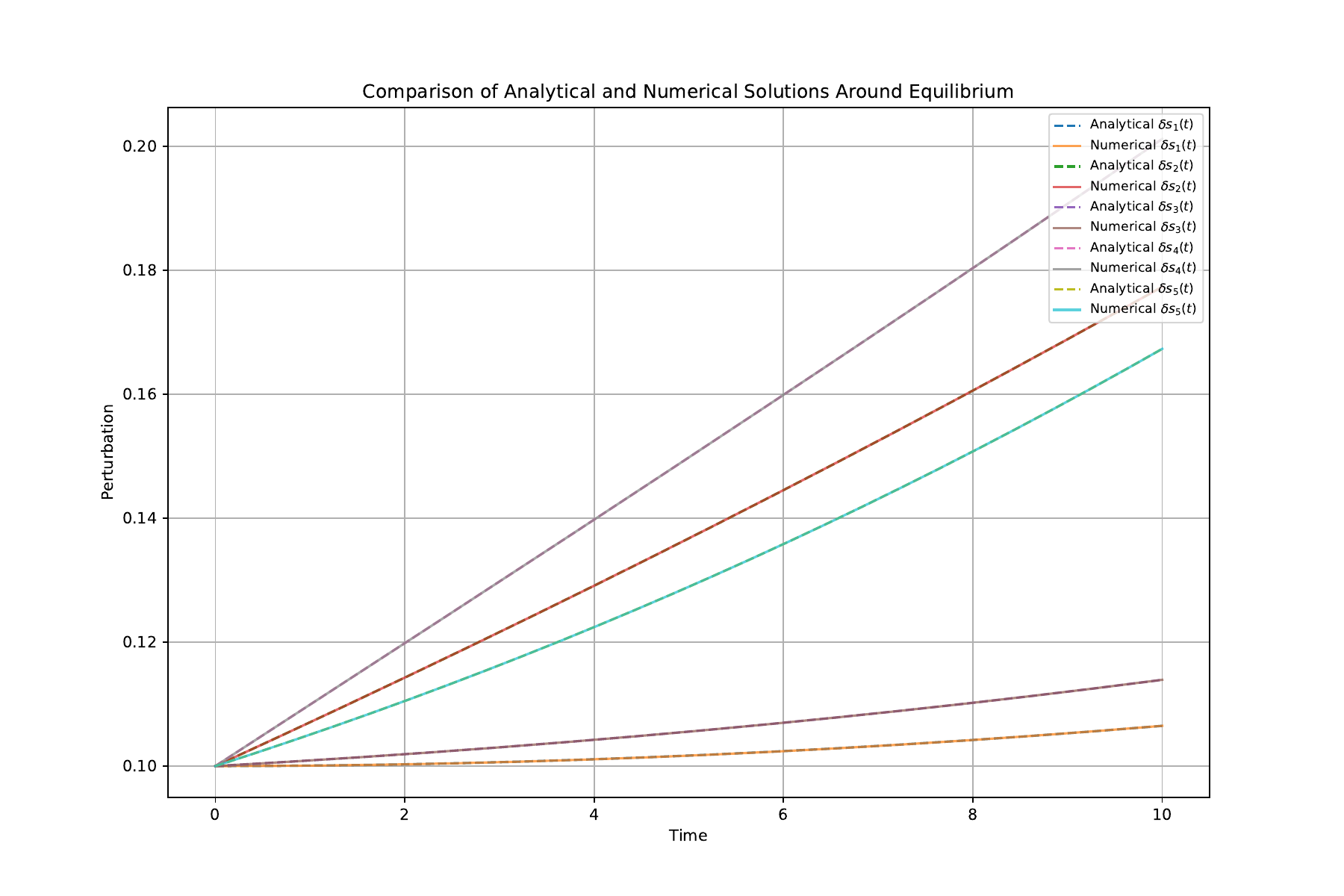}
    \caption{This figure depicts the analytical and numerical solutions of this linearized system of equations around equilibrium.}
    \label{fig6}
\end{figure}

Since, analytical solutions are not feasible for time dependent dynamics, we solved the system numerically using the Runge Kutta $4$th order method implemented in Python. The system was solved over a $24$hr period and cortisol levels were extracted for analysis. Figure(\ref{fig6}) compares the analytical and numerical solutions of the system for normal condition. The comparison of analytical and numerical solutions serves as a validation benchmark for model fidelity. While full numerical simulations are required for simulating long-term, non linear dynamics under time varying stress or feedback, the close match between the linearized and numerical solutions around equilibrium reinforces confidence in the model structure. \\

Although the original nonlinear system captures the full dynamics of the hormonal and glandular interactions in response to stress, its analytical intractability due to the coupling and nonlinearity introduced by feedback functions and time varying inputs, motivates the use of linearized approximations near steady state regimes. By determining equilibrium points under constant stress input and then applying first order perturbation expansions, the linearized system becomes a powerful tool for analyzing the local stability and response characteristics of the HPA axis.
\section*{Data availability}
The data that support the findings of this study are available from
mamta211935@cuh.ac.in. The relevant code used for the modeling will
also be provided to readers upon reasonable request.
%\section{References}
%\bibliographystyle{ieeetr}
\bibliography{refss}
\end{document}